%% file: main.tex
\documentclass[journal,10pt]{IEEEtran}
\IEEEoverridecommandlockouts
\usepackage{cite}
\usepackage{amsmath,amssymb,amsfonts,amsthm}
\usepackage{graphicx}
\usepackage[T1]{fontenc}
\usepackage{xcolor}
\usepackage{stackengine}
\usepackage{verbatim}
\usepackage{cite}

\usepackage{epstopdf}
\usepackage{enumerate}

\usepackage{color}
\usepackage{url}
\usepackage{balance}

\usepackage[caption=false]{subfig}
\usepackage{floatrow}

\usepackage{tikz}
\input{symbols.tex}

\begin{document}
\title{Beyond Diagonal RIS: Key to Next-Generation Integrated Sensing and Communications?}
\author{Tara Esmaeilbeig, Kumar Vijay Mishra and Mojtaba Soltanalian
\thanks{Tara Esmaeilbeig and Mojtaba Soltanalian are with the ECE Departement, University of Illinois at Chicago, Chicago, IL 60607 USA. Email: \{zesmae2, msol\}@uic.edu. Kumar Vijay Mishra is with the United States DEVCOM Army Research Laboratory, Adelphi, MD 20783 USA. E-mail: kvm@ieee.org.}
\thanks{This work was sponsored by the Army Research Office, accomplished under Grant Number W911NF-22-1-0263. The views and conclusions contained in this document are those of the authors and should not be interpreted as representing the official policies, either expressed or implied, of the Army Research Office or the U.S. Government. The U.S. Government is authorized to reproduce and distribute reprints for
Government purposes notwithstanding any copyright notation herein.}
}
\maketitle
\begin{abstract}
Reconfigurable intelligent surface (RIS) have introduced unprecedented flexibility and adaptability toward smart wireless channels. Recent research on integrated sensing and communication (ISAC) systems has demonstrated that RIS platforms enable enhanced signal quality, coverage, and link capacity. In this paper, we explore the application of fully-connected beyond diagonal RIS (BD-RIS) to ISAC systems. BD-RIS introduces additional degrees of freedom by allowing non-zero off-diagonal elements for the scattering matrix, potentially enabling further functionalities and performance enhancements. In particular, we consider the joint design objective of maximizing the weighted  sum of the signal-to-noise ratio (SNR) at the radar receiver and communication users by leveraging the extra degrees-of-freedom offered in the BD-RIS setting. These degrees-of-freedom are unleashed by formulating an alternating optimization process over known and auxiliary (latent) variables of such systems. Our numerical results reveal the
advantages of deploying BD-RIS in the context of ISAC and the effectiveness of
the proposed algorithm by improving the SNR values for both  radar and communication users by several orders of magnitude. 
\end{abstract}


\section{Introduction}\label{sec:intro} 
Reconfigurable intelligent surfaces (RIS) have  demonstrated  great  potential in realizing smart programmable environments  by modifying  wireless channels \cite{hodge2023index,hodge2020intelligent} and establishing  reliable links  for  communication and sensing~\cite{chepuri2023integrated}. In particular, RIS have been designed  to enhance  energy efficiency~\cite{huang2019reconfigurable}, physical  security~\cite{shi2024secrecy,mishra2022optm3sec}, radar target estimation and detection~\cite{esmaeilbeig2022irs}, and facilitate integrated sensing and communications (ISAC)~\cite{Shtaiwi2023sum,Liu2022joint}. In this paper, we focus on RIS-enabled ISAC.

 ISAC improves spectrum access and usage while reducing both hardware, power, and signaling costs \cite{mishra2019toward,liu2022,liu2022int}. The literature indicates that RIS benefits ISAC in  both line-of-sight (LoS) and non-LoS (NLoS) scenarios \cite{hu2024reconfigurable}. When LoS links exist between base station (BS) to communications users or radar targets, RIS is used to compensate  for the propagation loss~\cite{liu2023snr,wei2023multi}. When the LoS to either users or targets is blocked, RIS is utilized to  establish a virtual LoS or NLoS link to bypass  the obstructions~\cite{Esmaeilbeig2023Quantized,hua2022joint}. 

All of the above-mentioned  works employ conventional RIS architecture characterized by a diagonal scattering matrix (D-RIS), also
known as phase shift matrix. In D-RIS each element is controlled by a tunable impedance connected to ground. Whereas in~\cite{shen2022modeling}, beyond  diagonal RIS (BD-RIS) is introduced by connecting the elements of RIS to each other through a reconfigurable impedance network. When all or a group of the elements are connected to each other the  BD-RIS is called fully-connected or group-connected BD-RIS, respectively. The maximum  degrees-of-freedom  offered by the  fully-connected BD-RIS  has shown great potential in meeting  the upper bound of  received signal power in~\cite{nerini2023closed}. BD-RIS extends the concept of conventional RIS by actively manipulating the cross-polarization components, inter-element coupling, and coherent scattering effects \cite{li2023beyond,wu2023intelligent}. By dynamically adjusting these off-diagonal elements, beyond-diagonal RIS reshapes the wavefront in ways previously unattainable, empowering ISAC systems with enhanced resolution, interference suppression, and target discrimination capabilities~\cite{fang2023low,wang2023dual,santamaria2023snr}. In this paper, we  explore  the  benefit of fully-connected BD-RIS in ISAC applications. In general,  ISAC  systems may follow communications-centric design, radar-centric or integrated design~\cite{zhang2021overview}. Prior works, including~\cite{wang2023dual}, have focused on radar-centric design of BD-RIS for ISAC by maximizing  the target detection probability subject to  a minimum quality of service at communications users. 
 
 Contrary to these studies, we employ  a joint  communications and sensing design that offers a tunable trade-off between communications and  sensing performance. In particular, our design criterion is  a weighted sum of the  signal-to-noise ratio (SNR) at the  communications and  radar (sensing) receivers. This approach of weighted joint criterion has previously been considered in ISAC system design studies \cite{liu2020co,Chiriyath2017radar}. We show that optimization of this objective with respect to the hardware constraints of fully-connected BD-RIS is nonconvex. In order to tackle the resulting complicated non-convex optimization problem, we develop an efficient alternating optimization algorithm following a penalty-based framework for unimodular  quartic  optimization  over a finite alphabet set. Our numerical simulations, study the  performance of BD-RIS with different number of resolution  bits. Additionally, we  illustrate that the BD-RIS  outperforms the D-RIS with the same number of reflecting elements in terms of enhancing  the  communications  SNR, radar SNR, and, hence, their weighted sum.  

Throughout this paper, we use bold lowercase and bold uppercase letters for vectors and matrices, respectively. The identity matrix of  size $s\times s$ is denoted by $\mbI_s$. The sets of complex and real numbers are $\mathbb{C}$ and $\mathbb{R}$, respectively;  $(\cdot)^{\top}$ and $(\cdot)^{\mathrm{H}}$ are the vector/matrix transpose and the Hermitian transpose, respectively; trace of a matrix is  $\operatorname{Tr}(.)$. The Kronecker product is  denoted by $\otimes$. The vectorized form of a matrix $\mbB$ is written as $\vec{\mbB}$. 
The  maximum eigenvalue of $\mbB$ is  denoted by $\lambda_{max}(\mbB)$. The  angle/phase and real part of a complex number is denoted by $\arg{\cdot}$ and $\Re{\cdot}$, respectively. The operation $\mathrm{vec}_{_{K,L}}^{-1}\left(\mbc\right)$ reshapes the input vector $\mbc\in\mathbb{C}^{KL}$ into a  matrix $\mbC\in\mathbb{C}^{K \times L}$ such that $\vec{\mbC}=c$. For a symmetric $n \times n$ matrix $\mbA$, the half-vectorization, denoted by $\operatorname{vech}(\mbA)$, is the $(n(n + 1)/2) \times 1$ column vector obtained by vectorizing only the lower triangular part of $\mbA$ i.e. 
$\operatorname{vech}(\mbA)= [\mbA_{1,1}, \ldots, \mbA_{n, 1}, \mbA_{2,2}, \ldots, \mbA_{n, 2}, \ldots, \mbA_{n-1, n-1}, \mbA_{n, n-1},$ $\mbA_{n, n}]^{\mathrm{T}}$. Moreover, the unique \emph{duplication} matrix $\mbD_n$ transforms $\operatorname{vech}(\mbA)$ into  $\operatorname{vec}(\mbA)$ as $\operatorname{vec}(\mbA)=\mbD_n \operatorname{vech}(\mbA)$.

\section{System Model}\label{sec::model}
\begin{figure}[t]
\centering
	\input{fig1.tex}
     \caption{A illustration of BD-RIS-enabled ISAC system. When the LoS is blocked, the NLoS paths via the BD-RIS allow for establishing the link between the targets/users with the DFBS. 
     }
\label{fig_1}
\end{figure}
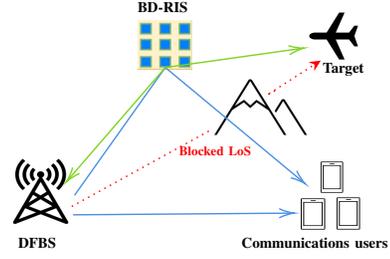

Consider a BD-RIS-enabled  ISAC system (Fig.~\ref{fig_1}) comprising a dual-function base station (DFBS) with $N$ elements each in transmit (Tx)  and receive (Rx) antenna arrays. The BD-RIS comprises $L=L_x \times L_y$  reflecting elements arranged as a uniform planar array (UPA) with $L_x$ ($L_y$) elements along the x- (y-) axes in the Cartesian coordinate plane. Define the RIS steering vector 
$\mba(\theta_h,\theta_v)=\mba_x(\theta_h,\theta_v) \otimes \mba_y(\theta_h,\theta_v)$, where $\theta_h$ ($\theta_v$) is the azimuth (elevation) angle, $\mba_x(\theta_h,\theta_v)=[1,e^{\textrm{j}\frac{2\pi d}{\lambda} cos \theta_h sin \theta_v},\ldots,e^{\textrm{j}\frac{2\pi d (L_x-1)}{\lambda} cos \theta_h sin \theta_v}]^{\top}$ and  $\mba_y(\theta_h,\theta_v)=[1,e^{\textrm{j}\frac{2\pi d}{\lambda} cos \theta_h sin \theta_v},\ldots,e^{\textrm{j}\frac{2\pi d (L_y-1)}{\lambda} cos \theta_h sin \theta_v}]^{\top}$, $\lambda=c/f_c$ is the carrier wavelength, $c=3 \times 10^8$ m/s is the speed of light, $f_c$ is the carrier frequency, and $d=0.5\lambda$ is the  inter-element spacing. The fully-connected BD-RIS operation is characterized by the scattering or phase matrix $\bPhi$ which is a  complex symmetric unitary, i.e.,   \par\noindent\small
 \begin{align} \label{bd-ris1}
 \bPhi=\bPhi^{\T} &\text{and} \;\; \bPhi^{\H}\bPhi=\mbI.
\end{align} \normalsize

The phases may be assumed continuous or discrete/quantized. In general, discrete phase settings simplify the hardware implementation and control/programming of the RIS. Since discrete phase states are achieved using digital switches or phase shifters, power consumption is lower than continuously variable analog counterparts. 
For discrete-valued phases  with  resolution $M$,  we  have $[\bPhi]_{_{kl}} \in \bOmega_{_M}$, where \par\noindent\small
\begin{align} \label{eq:3}
\bOmega_{_M}&= \left\{ s= e^{\textrm{j}\omega}, \omega \in \Pi_{_M}\right\}, \nonumber\\
\Pi_{_M}&=\left\{1,\frac{2\pi}{M},\cdots,\frac{2\pi(M-1)}{M}\right\}.
\end{align}\normalsize
The continuous  BD-RIS is  equivalent to the discrete RIS with infinite   resolution i.e. $M \rightarrow \infty$~\cite{nerini2023discrete}. In other words, we have $[\bPhi]_{_{kl}} \in \bOmega$ where 
$\bOmega= \left\{ s= e^{\textrm{j}\omega}, \omega \in [0,2\pi)\right\}$.
In this paper, to achieve a discrete-valued BD-RIS phase design, we obtain the optimal phases on the set $\bOmega$ and then project the  solution onto the set $\bOmega_M$.

We now describe the  received communications signal at the  users and the sensing  signal  back-scattered from the target and received at the DFBS. \\
\noindent\textbf{Communications Rx signal:} Denote the direct  or LoS 
channel  state information (CSI) and RIS-reflected non-line-of-sight (NLoS) CSI matrices by $\mbF \in \complexC{K \times N}$ and $\mbH \in \complexC{K \times L}$, respectively. Then, at each communications receiver after  sampling, 
 the discrete-time received signal is $\mby_{c,k}$. Concatenating the received signals from all  users, we obtain the  $K \times 1$ vector:\par\noindent\small 
\begin{equation}
\mby_c=(\mbF+\mbH\bPhi\mbG) \mbP \mbs +  \mbn_c,
\end{equation}\normalsize
where  $\mbP\in \complexC{N \times K}$ is the DFBS precoder, the  Tx-RIS CSI matrix $\mbG \in \complexC{L \times N}$ is assumed to be estimated \textit{a priori} through suitable channel estimation techniques and  $\mbn_c \sim \mathcal{N}(\bzero,\sigma^2_{c}\mbI_{K})$  is  the spatially and temporally white noise  at communications receivers. \\
\noindent\textbf{Sensing Rx signal at DFBS:} Consider a Swerling-0 \cite{skolnik2008radar} radar target located at a  NLoS location with respect to the DFBS. The 
direction-of-arrival (DoA) is $(\theta_{h_t},\theta_{v_t})$ with respect to the RIS and radar cross-section (RCS) $\alpha_{_T}$. Define $\mbR=\alpha_{_T} \mbG^{T}\bPhi \mba(\theta_{h_t},\theta_{v_t}) \mba(\theta_{h_t},\theta_{v_t})^{\T}\bPhi \mbG$. 

Stacking the echoes for all receiver antennas, the $N \times 1$  
signal at radar receiver after down conversion and  sampling is\par\noindent\small
\begin{equation}\mby_{_r}
= \mbR\mbP \mbs +\mbn_r, 
\end{equation}\normalsize
where $\mbn_r \sim \mathcal{N}(\bzero,\sigma^2_{r}\mbI_{N})$  is  the noise  at radar receiver.

Denote  $\mbC=\mbF+\mbH\bPhi\mbG$.
The output SNR at communications and DFBS Rx 
are, respectively,\par\noindent\small
\begin{align}
 \textrm{SNR}_{c}&=\frac{1}{\sigma^2_{c}}\Tr{\mbC \mbP \mbP^{\H}\mbC^{\H}},  
  \textrm{and }\label{snr_c}\\
  \textrm{SNR}_{r}&=\frac{1}{\sigma^2_{r}}\Tr{\mbR \mbP \mbP^{\H}\mbR^{\H}}. \label{snr_r}
\end{align}\normalsize
In the next section, we use~\eqref{snr_c}-\eqref{snr_r} to formalize the  design problem for BD-RIS-enabled ISAC  system.

\section{BD-RIS design for ISAC}
As the design criterion under BD-RIS hardware constraints \eqref{bd-ris1} and~\eqref{eq:3}, we use the weighted sum of the radar and communications SNRs, with a weight factor $\beta$ as $\textrm{SNR}_{_T}=\beta \textrm{SNR}_{r} + (1-\beta) \textrm{SNR}_{c}$. Our design problem is \par\noindent\small
\begin{align}\label{eq:opt1}
\mathcal{P}_{1}:&\;\underset{\bPhi \in \bOmega_{M}^{^{L \times L}}}{\textrm{maximize}} \; \; \textrm{SNR}_{_T} \nonumber \\
&\;\textrm{subject to} \; \;
\bPhi^{\H}\bPhi=\mbI \;\; \text{and} \;\;\; \bPhi=\bPhi^{\T}.
\end{align} \normalsize
Since $\bPhi$ is symmetric, we solve $\calP_1$ w.r.t  $\bphi=\operatorname{vech}(\bPhi) \in \bOmega_{M}^{^{L(L+1)/2}}$ by means of a penalty-based optimization  method. Following Lemmata 1 and 2, state that the SNR at  communications  and the radar receivers are  quadratic and  quartic with respect to $\bphi$, respectively.
\begin{lemma}
For a symmetric scattering matrix $\bPhi$, 
 \begin{align}\label{eq:1}
\text{SNR}_{c}&= \bphi^{\H} \widetilde{\mbQ} \bphi +2\Re{\mbq^{\H}\bphi}+ \mbf^{H}\widehat{\mbP}\mbf,
\end{align}\normalsize
where $\widetilde{\mbQ}=\mbD_{_L}^{\T}(\mbG^{H}\otimes \mbH)^{\H}\widehat{\mbP} (\mbG^{H}\otimes \mbH)\mbD_{_L}$, with $\mbD_{_L}$ denoting the duplication matrix for $\bPhi$, and $\widehat{\mbP}=\frac{1}{\sigma^2_{c}} \left(\mbP^{\T}\otimes \mbI_{K}\right)^{\H}\left(\mbP^{\T}\otimes \mbI_{K}\right)$. Also, $\mbq=\mbD_{_L}^{\T}(\mbG^{H}\otimes \mbH)^{\H} \widehat{\mbP} \mbf$.
\end{lemma}
\begin{IEEEproof} From~\eqref{snr_c} we have\par\noindent\small
\begin{align} \label{eq:SNRc}
\textrm{SNR}_{c}&=\frac{1}{\sigma^2_{c}}\Tr{\mbC \mbP \mbP^{\H}\mbC^{\H}}=\frac{1}{\sigma^2_{c}} \Vec{\mbC\mbP}^{\H}\Vec{\mbC\mbP} \nonumber \\
&=\frac{1}{\sigma^2_{c}}\Vec{\mbC}^{\H}\left(\mbP^{\T}\otimes \mbI_{K}\right)^{\H}\left(\mbP^{\T}\otimes \mbI_{K}\right)\Vec{\mbC} \nonumber \\
&= \Vec{\mbC}^{\H}\widehat{\mbP} \Vec{\mbC} \nonumber \\
&= \mbf^{H}\mbQ\mbf +\mbf^{H} \widehat{\mbP} (\mbG^{H}\otimes \mbH) \bphi +\bphi^{\H} (\mbG^{H}\otimes \mbH)^{\H} \widehat{\mbP} \mbf  \nonumber \\
&+ \bphi^{\H} (\mbG^{H}\otimes \mbH)^{\H} \widehat{\mbP} (\mbG^{H}\otimes \mbH) \bphi,
\end{align}\normalsize
where the last  equality  used  $\Vec{\mbC}=\Vec{\mbF}+(\mbG^{H}\otimes \mbH)\Vec{\bPhi}=\mbf+(\mbG^{H}\otimes \mbH)\mbD_{_L}\bphi$. Substituting  $\widetilde{\mbQ}$ and $\mbq$  in~\eqref{eq:SNRc} yields \eqref{eq:1}. 
\end{IEEEproof}
\begin{lemma} For a symmetric scattering matrix $\bPhi$,
   \par\noindent\small
 \begin{align}\label{eq:2}
\text{SNR}_{r}&=\bphi^{\H} \bar{\bQ}(\bPhi) \bphi,
\end{align}\normalsize\par\noindent\small
where\begin{align}
 \bar{\bQ}(\bPhi)&=\mbD_{_L}^{\T}\left(\mbG^{\T}\otimes \mbG^{\T}\bPhi\mba(\theta_{h_t},\theta_{v_t})\mba^{\T}(\theta_{h_t},\theta_{v_t})\right)^{\H} \bar{\mbP} \nonumber \\
& \left(\mbG^{\T}\otimes \mbG^{\T}\bPhi\mba(\theta_{h_t},\theta_{v_t})\mba^{\T}(\theta_{h_t},\theta_{v_t})\right)\mbD_{_L}, \label{eq:14} \\
\bar{\mbP}&=\frac{|\alpha_{T}|^{2}}{\sigma^2_{r}}\left(\mbP^{\T}\otimes \mbI_{N}\right)^{\H}
\left(\mbP^{\T}\otimes \mbI_{N}\right). 
\end{align}\normalsize
\end{lemma}
\begin{IEEEproof} 
From~\eqref{snr_r} we have\par\noindent\small
\begin{align}
\textrm{SNR}_{r}&=\frac{1}{\sigma^2_{r}}\Tr{\mbR \mbP \mbP^{\H}\mbR^{\H}}=\frac{1}{\sigma^2_{r}} \Vec{\mbR\mbP}^{\H}\Vec{\mbR\mbP} \nonumber \\
&=\frac{1}{\sigma^2_{r}}\Vec{\mbR}^{\H}\left(\mbP^{\T}\otimes \mbI_{N}\right)^{\H}\left(\mbP^{\T}\otimes \mbI_{N}\right)\Vec{\mbR} \nonumber \\
&= \bphi^{\H} \bar{\bQ}(\bPhi) \bphi,
\end{align}\normalsize
where the  last equality used 
$\Vec{\mbR} 
= \alpha_{T} \Vec{\mbG^{\T}\bPhi\mba(\theta_{h_t},\theta_{v_t})\mba^{\T}(\theta_{h_t},\theta_{v_t})\bPhi \mbG} 
=\alpha_{T}\left(\mbG^{\T}\otimes \mbG^{\T}\bPhi\mba(\theta_{h_t},\theta_{v_t})\mba^{\T}(\theta_{h_t},\theta_{v_t})\right)\mbD_{_L}\bphi$. 
\end{IEEEproof}
 From Lemma 1-2, we conclude that  the  objective  function in $\calP_1$ is  quartic  w.r.t  $\bPhi$. Define
$f(\bphi) = \textrm{SNR}_{_T} = \beta \bphi^{\H} \bar{\bQ}(\bphi) \bphi + (1-\beta) \bphi^{\H} \widetilde{\mbQ} \bphi 
+2(1-\beta) \Re{\mbq^{\H}\bphi }$.
In order to  reduce the  problem from  quartic to  quadratic optimization we introduce a latent (auxiliary) variable  $\bPhi_{0} \in \bOmega^{L \times L}$ as a symmetric  matrix. Define  the new objective as\par\noindent\small
\begin{align}\label{eq:19}
f(\bphi,\bphi_0)&= \beta \bphi^{\H} \bar{\bQ}(\bPhi_0) \bphi + (1-\beta) \bphi^{\H} \widetilde{\mbQ} \bphi \nonumber \\
&+2(1-\beta) \Re{\mbq^{\H}\bphi }.
\end{align}\normalsize
In order to have the  surrogate objective equal to the  original  objective, i.e. $f(\bphi,\bphi_0)=f(\bphi)$, we propose to optimize $f(\bphi,\bphi_0)$ under the  constraint  $\bphi=\bphi_0$, where $\boldsymbol{\phi}_0=\vech{\bPhi_0}$.

Additionally, in order to impose the  constraint $\bPhi \in \bOmega_{_M}^{^{L \times L}}$ in~\eqref{eq:opt1}, we introduce a  second latent variable, $\bPhi_{1} \in \bOmega^{^{L \times L}}_{_M}$ as a symmetric matrix and impose $\bPhi=\bPhi_1$. Denote $\boldsymbol{\phi}_1=\vech{\bPhi_1}$. 
In order to impose the  unitarity constraint on  $\bPhi$ in $\calP_1$, we introduce a third  auxiliary variable $\mbU$ with the constraints  $\mbU=\bPhi$ and $\mbU^{\H}\mbU=\mbI$. Introducing these auxiliary variables into problem $\calP_1$, we obtain the  equivalent problem \par\noindent\small
\begin{align}\label{eq:opt2}
\mathcal{P}_{2}:\;\underset{\bphi,\bphi_{0},\bphi_{1},\mbU}{\textrm{maximize}} &\; \;  f(\bphi,\bphi_0)\nonumber \\
\;\textrm{subject to} &\; \bphi=\bphi_0=\bphi_1=\vech{\mbU},\\
& \mbU^{\H}\mbU=\mbI, \; \label{eq:200}\\
& \; \bphi,\bphi_{0} \in \bOmega^{^{L(L+1)/2}}  \text{and} \; \bphi_{1} \in \bOmega^{^{L(L+1)/2}}_{M}.\label{eq:199}
\end{align}\normalsize
We solve this non-convex problem  
using the framework proposed in~\cite{yu2020quadratic} for  unimodular quadratic optimization.
\section{Optimization Framework}
The  Lagrangian  of  $\calP_2$ is\par\noindent\small
\begin{align}\label{eq:21}
\calL(\bphi,\bphi_{0},\bphi_{1},\mbU&)=-f(\bphi,\bphi_0)
+\sum_{i=0}^{1} \frac{\rho_i}{2}\|\bphi_i-\bphi\|^2_2  \\
&+\frac{\rho_{_2}}{2}\|\vec{\mbU}-\bphi\|^2_2  
, \nonumber
\end{align}\normalsize
where  $\rho_i >0$, $i \in \{0,\ldots,3\}$ are  penalty parameters. Next, we employ a  penalty based  framework and  determine $(\bphi,\bphi_{0},\bphi_{1},\mbU)$ in an alternating iterative  fashion to minimize  $\calL(\bphi,\bphi_{0},\bphi_{1},\mbU)$ under the constraints in~\eqref{eq:200}-\eqref{eq:199}.\\

\textbf{Update of $\bphi$:} With fixed  $\bphi^{(r)}_{0},\bphi^{(r)}_{1},\mbU^{(r)}$  at iteration r we need to minimize  $\calL(\bphi,\bphi^{(r)}_{0},\bphi^{(r)}_{1},\mbU^{(r)})$ w.r.t $\bphi$, which from~\eqref{eq:19}-\eqref{eq:21}, is equivalent to \par\noindent\small
\begin{align}\label{eq:opt-phi}
&\underset{\bphi\in \bOmega^{^{L(L+1)/2}}}{\textrm{minimize}} \; \;  \bphi^{\H} \mbR^{(r)}\bphi +\Re{\mbc^{(r)\H}\bphi}
\end{align}\normalsize
where   $\mbR^{(r)}=-\beta \bar{\mbQ}(\bphi_0^{(r)})+(\beta-1)\widetilde{\mbQ}$ and $\mbc^{(r)}=2(\beta-1)\mbq -\sum_{i=0}^{1} \rho_i\bphi_i^{(r-1)}-\rho_{_2}\vec{\mbU}$.~\eqref{eq:opt-phi} is equivalent to\par\noindent\small
\begin{equation}\label{neg13}
  \underset{\bar \bphi\in\bOmega^{^{L(L+1)/2+1}}}
{\textrm{minimize}} \quad
\bar\bphi^{\mathrm{H}}
\mbA^{(r)}
 \bar\bphi,   
 \end{equation}\normalsize
where $\mbA^{(r)}=\begin{bmatrix} \mbR^{(r)} &\mbc^{(r)}/2\\ \mbc^{(r)\H}/2& 0 \end{bmatrix}$ and $\bar \bphi=[\bphi^{\T} \; 1]^{\T}$. This problem with feasible set $\bOmega^{^{L(L+1)/2+1}}$ is a unimodular quadratic program for which the following iterations will  lead to non-increasing objective\par\noindent\small
\begin{equation}\label{phi_update}
\bar\bphi^{(r,s+1)}=e^{\j \arg{\bar\mbA^{(r)} \bar\bphi^{(r,s)}}}.
\end{equation}\normalsize
Note that $s$ denotes the iteration number over~\eqref{phi_update} and we used the diagonal loading technique, i.e., $\bar\mbA^{(r)}  = \lambda_{m}\mbI-\mbA^{(r)}$, where the loading parameter
$\lambda_{m} \geq \lambda_{max}(\mbA^{(r)})$~\cite{soltanalian2014designing}.\\

\textbf{Update of $\bphi_0$:} The  Lagrangian  function $\calL(\bphi^{(r)},\bphi_{0},\bphi_{1}^{(r-1)},\mbU^{(r-1)})$  is quadratic w.r.t  $\bphi_0$. To show this, from~\eqref{eq:14} we write 
$\bphi^{(r)\H} \bar{\bQ}(\bPhi_0) \bphi^{(r)}= \bphi_0^{\H} \bar{\bar{\mbQ}}(\bPhi^{(r)}) \bphi_0$,
where \par \noindent \small
\begin{equation}\label{eq:26}
 \bar{\bar{\mbQ}}(\bphi^{(r)})=  \frac{|\alpha_{T}|^{2}}{\sigma^2_{r}}\left(\mbG^{\T}\bPhi^{(r)\T}\mba\mba^{\T}\otimes \mbG^{\T}\right)^{\H}
\bar{\mbP}\left( \mbG^{\T}\bPhi^{(r)\T}\mba\mba^{\T}\otimes \mbG^{\T}\right) 
\end{equation} \normalsize
and  $\bPhi^{(r)}=\mathrm{vec}_{_{L,L}}^{-1}\left(\mbD_{_L}\bphi^{(r)}\right)$. For the  sake of brevity,  we denoted $\mba(\theta_{h_t},\theta_{v_t})$ by $\mba$.
By  substituting~\eqref{eq:26} in~\eqref{eq:21}, the  minimization of the  Lagrangian w.r.t  $\bphi_0$ is  equivalent to \par\noindent\small
\begin{align}\label{eq:opt-phi_0}
&\underset{\bphi_0 \in \bOmega^{^{L(L+1)/2}}}{\textrm{minimize}} \; \;  \bphi_0^{\H} \bar{\mbR}^{(r)}\bphi_0 +\Re{\bar{\mbc}^{(r)\H}\bphi_0},
\end{align}\normalsize
where $\bar{\mbR}^{(r)}=-\beta  \bar{\bar{\mbQ}}(\bphi^{(r)}) $ and $\bar{\mbc}^{(r)}= -\rho_0\bphi^{(r)}$. Similar to the update  for  $\bphi$,  by letting $\hat\mbA^{(r)}=\begin{bmatrix} \bar \mbR^{(r)} &\bar \mbc^{(r)}/2\\ \bar \mbc^{(r)\H}/2& 0 \end{bmatrix}$ and $\bar \bphi_0=[\bphi_0^{\T} \; 1]^{\T}$, similar  power method like iterations as in~\eqref{phi_update}, will be applicable  to the problem. We have  \par\noindent\small
\begin{equation}\label{phi_0_update}
\bar \bphi_0^{(r,t+1)}=e^{\j \arg{\widehat{\widehat{\mbA}}^{(r)} \bar\bphi_0^{(r,t)}}}.
\end{equation}\normalsize
where  $t$ is the  number for  power method like  iterations and  $\widehat{\widehat{\mbA}}^{(r)}  = \lambda_{m}\mbI-\widehat{\mbA}^{(r)}$, with the loading parameter is chosen such that
$\lambda_{m} \geq \lambda_{max}(\widehat{\mbA}^{(r)})$.\\
\textbf{Update of $\bphi_1$:}    
Given  $\bphi^{(r)},\bphi^{(r)}_{0},\mbU^{(r-1)}$,  we minimize 
$\calL(\bphi^{(r)},\bphi^{(r)}_{0},\bphi_{1},\mbU^{(r-1)})$ under the  finite alphabet constraint  $\bphi_1 \in \bOmega_M
^{L^{2} \times 1}$. Analogously, we solve \par\noindent\small
\begin{align}\label{eq:opt-phi_11}
&\underset{\bphi_1 \in \bOmega^{^{L(L+1)/2}}_{M}}{\textrm{maximize}} \; \;  \Re{\hat{\mbc}^{(r)\H}\bphi_1},
\end{align}\normalsize
where  $\hat{\mbc}^{(r)}=\rho_1\bphi^{(r)}$. Rewrite this problem as \par\noindent\small
\begin{align}\label{eq:opt-phi_12}
&\underset{\gamma_l}{\textrm{maximize}} \; \;  \cos{(\gamma_l-\bpsi_l)}\nonumber \\
&\;\textrm{subject to} \; \;
\gamma_l \in  \left\{1,\frac{2\pi}{M},\cdots,\frac{2\pi(M-1)}{M}\right\} \;\;, 
\end{align}\normalsize
where $\gamma_l$ is the  phase of $\bphi_{1,l}$ and  $\bpsi_l $ is  the phase of  $\hat{\mbc}^{(r)}_l$. The solution  is  obtained by an  exhaustive search  for  $m$ such that 
$m^{(r)}_l 
=\underset{m}{\textrm{argmax}} \; \;  \cos{(2\pi \frac{m}{M}-\bpsi_l)}$ 
$m \in \{0,\ldots,M-1\}$. 
Then, for $l \in \{1,\ldots,L^{2}\}$, the  update of the $l$-th element in $\bphi_1^{(r)}$ is \par\noindent\small
\begin{equation}\label{update_phi_1}
\bphi_{1_l}^{(r)}= e^{\j2\pi \frac{m^{(r)}_l}{M}}.
\end{equation}\normalsize\\
\textbf{Update of $\mbU$:} 
Consider  the following problem \par\noindent\small
\begin{align}\label{eq:opt32}
\mathcal{P}_{2}:&\;\underset{\mbU}{\textrm{minimize}} \; \;  \|\bPhi^{(r)}-\mbU\|_{_2}\nonumber \\
&\;\textrm{subject to} \; \;
\mbU^{\H}\mbU=\mbI .
\end{align}\normalsize
Consider  the singular value decomposition (SVD) of $\bPhi^{(r)}$ as 
$\bPhi^{(r)}=\mbU_1 \Sigma \mbU_2^{\H}$,
then the solution of $\calP_2$ is  given by~\cite{he2009designing} \par\noindent\small
\begin{equation} \label{update_U}
\mbU^{(r)}=\mbU_2 \mbU_1^{\H}.
\end{equation}\normalsize
Algorithm 1 summarizes the overall design procedure of  BD-RIS  scattering matrix.

\begin{algorithm}[h]
\caption{\normalsize Optimal design of BD-RIS scattering matrix for ISAC.}\label{alg:place}
    \begin{algorithmic}[1]
   \Statex \textbf{Input} Initializations   $\bphi^{(0)}$,  $\bphi_0^{(0)}$,  $\bphi_1^{(0)}$, $\mbU^{(0)}$, $r=1$.
   \State Set $s=1$, \textbf{while} $ \|\bar \bphi^{(r,s)}-\bar\bphi^{(r,s+1)}\| > \epsilon $ \textbf{do}  update  $\bar\bphi^{(r,s)}$ according to~\eqref{phi_update} and $s=s+1$.
   \State $\bphi^{(r)}=[\mbI_{_{L(L+1)/2}} \;\; \bzero]\bar \bphi^{(r,s)}$.
   \State Set $t=1$, \textbf{while} $ \|\bar\bphi_0^{(r,t)}-\bar\bphi_0^{(r,t+1)}\| > \epsilon_1 $ \textbf{do}  update  $\bar \bphi_0^{(r,t)}$ according to~\eqref{phi_0_update} and $t=t+1$.
   \State Update $\bphi_1^{(r)}$ and $\mbU^{(r)}$ as in~\eqref{update_phi_1} and~\eqref{update_U}, respectively.
   \State  $r=r+1$.
   \State  Repeat steps  $1$-$5$  until a prescribed stop criterion is satisfied, e.g., $ \|\bphi^{(r)}-\bphi^{(r+1)}\| \leq \epsilon_2 $.
   \Statex \textbf{Output} $\bPhi^{*}=\mathrm{vec}_{_{L,L}}^{-1}\left(\mbD_{_L}\bphi^{(r)}\right)$.
   \end{algorithmic}
\end{algorithm}
\begin{remark}
For continuous-valued scattering matrices i.e. $\bphi \in \bOmega^{L^2}$, it is  sufficient to remove  the latent variable  $\bphi_1$ from  $\calP_2$ and solve it with respect to $\bphi$, $\bphi_0$ and $\mbU$. \end{remark}
\begin{remark} 
Contrary to BD-RIS above, D-RIS design \cite{Esmaeilbeig2023Quantized} is less complex because it introduces a single latent variable to simplify quartic optimization into a quadratic alternative. 
\end{remark}
\section{Numerical Experiments}
\begin{figure}[t] 
    \centering
    \subfloat[][SNR$_{_T}$]{\includegraphics[width=0.49\textwidth ]{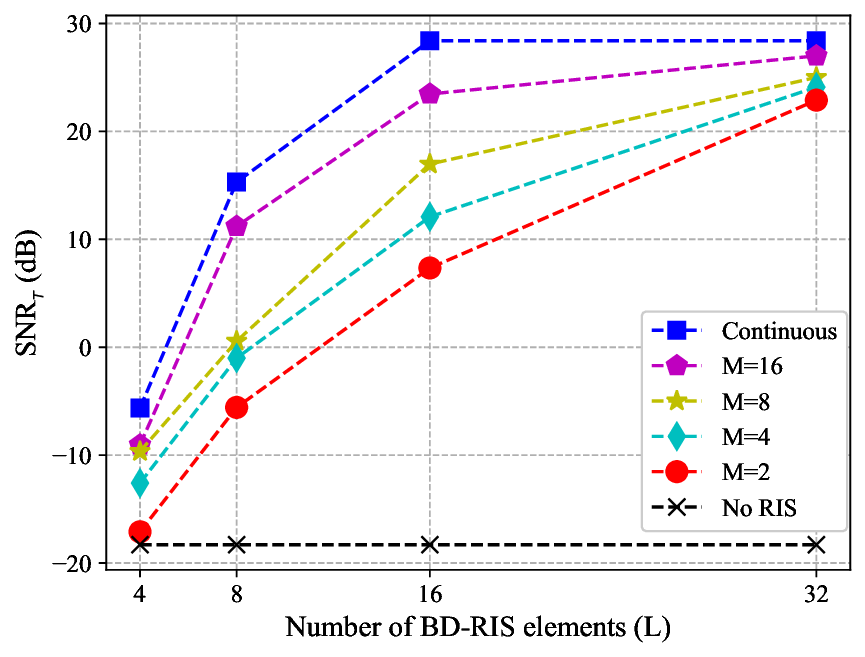}}
    \subfloat[][SNR$_c$ and SNR$_r$ vs. $\beta$]{\includegraphics[width=0.49\textwidth ]{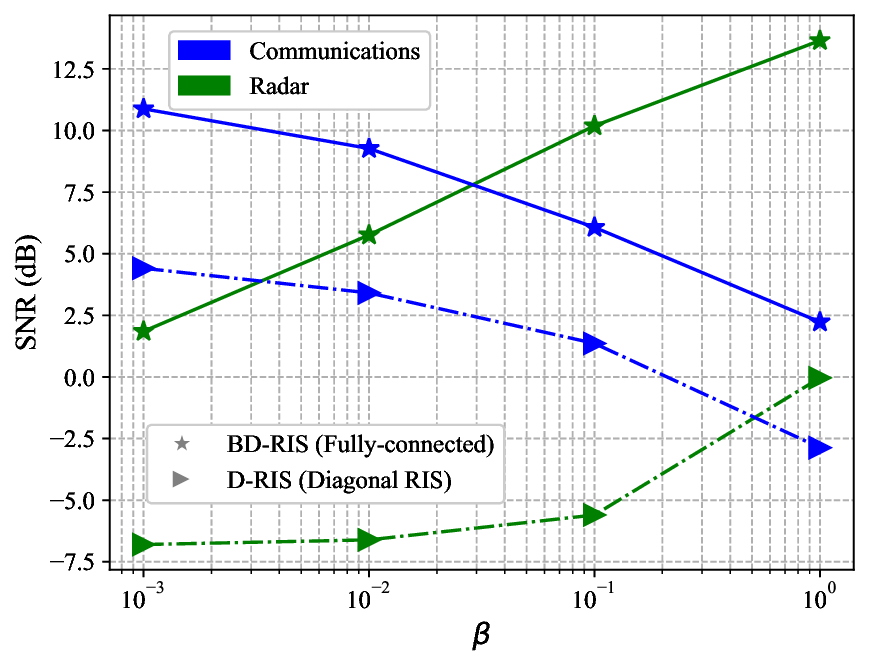}}
    \caption{(a) SNR$_{_T}$ is plotted against the increasing number of elements in BD-RIS. (b) Performance of D-RIS and  BD-RIS in terms of  SNR$_c$ and  SNR$_r$ vs.  different values of $\beta$.}\label{fig::4}
\end{figure}
To evaluate  our
proposed algorithm, we assume  the  DFBS located at $(0,0,0)$m, in  the  3-D Cartesian plane,  is equipped with $N = 8$ Tx and Rx antennas with half-wavelength
antenna spacing. The  ISAC system is serving  K = 4 single-antenna users located at $(20,20,0)$ m and simultaneously  detects a Swerling 0  point-like target  with complex reflectivity coefficient $\alpha_T \sim \mathcal{CN}(0,1)$ and located at $(20,20,0)$ m. The ISAC system is  being  assisted  by a BD-RIS located at $(30,30,0)$m. The variance of noise at  communication users and sensing   Rx is set at $\sigma^2_{c}=\sigma^2_{r}=-100$ dBm. We assumed an i.i.d  Rayleigh fading  channel in the DFBS to  BD-RIS, BD-RIS to users and  DFBS to users paths which we  abbreviate by RD,UR and UD,  respectively.    The path loss is modelled as $p_{ij}=p_0 d^{-\alpha_{ij}}$, where $p_0$ is the path loss at distance $1$m and $ij \in \{\text{RD, UR, UD}\}$.  We set $p_0=-30$ dB,  $\alpha_{_{\text{RD}}}=2$,   $\alpha_{_{\text{UD}}}=4$ and $\alpha_{_{\text{UR}}}=2.8$ for the paths that signal propagates through. The  CSI matrices are  generated according to $\mbG \sim \mathcal{CN}(0,p_{_{\text{RD}}}\mbI)$, $\mbH \sim \mathcal{CN}(0,p_{_{\text{UR}}}\mbI)$ and $\mbF \sim \mathcal{CN}(0,p_{_{\text{UD}}}\mbI)$. 

Fig.~\ref{fig::4}a illustrates the performance of Algorithm 1 for  different  resolutions of  BD-RIS with discrete valued optimized phases in comparison with the  continuous  phase and  no RIS. We set $\beta=0.5$,  and manually tuned the penalty parameters  $\rho_i$, $i \in \{0,\ldots,3\}$ until convergence.  To design BD-RIS with continuous phases, we adapt Algorithm 1 as per Remark 1. Fig.~\ref{fig::4}b shows that fully-connected BD-RIS outperforms conventional  D-RIS in terms of the optimized SNR$_r$ and SNR$_c$. Here, D-RIS is employed with  phases  optimized according to Remark 2. Moreover, changing $\beta$ in the interval $[10^{-3},1]$ tunes our  design criterion. Increasing $\beta$ shifts the system from communications-centric to radar-centric as evidenced from increasing (decreasing) SNR$_{r}$ (SNR$_c$).
\balance
\bibliographystyle{IEEEtran}
\bibliography{refs}
\end{document}

%% file: symbols.tex
\usepackage{amsmath}
\usepackage{amsthm}
\usepackage{dsfont}
\usepackage[noend]{algpseudocode}
\usepackage{algorithm}
\usepackage{seqsplit}
\usepackage{thmtools}
\usepackage{amssymb}

\def\bQ{\boldsymbol{Q}}

\def\bphi{\boldsymbol{\phi}}

\def\bPhi{\boldsymbol{\Phi}}

\def\bpsi{\boldsymbol{\psi}}

\def\bOmega{\boldsymbol{\Omega}}

\def\mba{\mathbf{a}}

\def\mbc{\mathbf{c}}

\def\mbf{\mathbf{f}}

\def\mbn{\mathbf{n}}

\def\mbq{\mathbf{q}}

\def\mbs{\mathbf{s}}

\def\mby{\mathbf{y}}

\def\mbA{\mathbf{A}}
\def\mbB{\mathbf{B}}
\def\mbC{\mathbf{C}}
\def\mbD{\mathbf{D}}

\def\mbF{\mathbf{F}}
\def\mbG{\mathbf{G}}
\def\mbH{\mathbf{H}}
\def\mbI{\mathbf{I}}

\def\mbP{\mathbf{P}}
\def\mbQ{\mathbf{Q}}
\def\mbR{\mathbf{R}}

\def\mbU{\mathbf{U}}

\def\calL{\mathcal{L}}

\def\calP{\mathcal{P}}

\def\bzero{\boldsymbol{0}}

\newcommand{\complexC}[1]{\mathds{C}^{#1}}

\def\Tr#1{\mathrm{Tr}\left(#1\right)}
\def\vec#1{\mathrm{vec}\left(#1\right)}
\def\vech#1{\mathrm{vech}\left(#1\right)}

\def\Re#1{\operatorname{Re}\left(#1\right)}

\def\arg#1{\operatorname{arg}\left(#1\right)}

\newtheorem{lemma}{Lemma}

\newtheorem{remark}{Remark}

\def\T{\top}
\def\H{\mathrm{H}}
\def\j{\mathrm{j}}

%% file: fig1.tex
\scalebox{0.7}{\tikzset{every picture/.style={line width=0.75pt}} 
\begin{tikzpicture}[x=0.75pt,y=0.75pt,yscale=-1,xscale=1]
\draw [color={rgb, 255:red, 255; green, 2; blue, 0 }  ,draw opacity=1 ] [dash pattern={on 0.84pt off 2.51pt}]  (264,200) -- (373.43,139.78) ;
\draw (400,128.78) node  {\includegraphics[width=52.5pt,height=52.5pt]{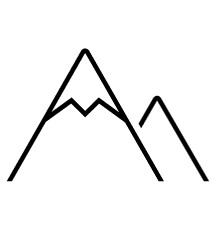}};
\draw (242.43,188.78) node  {\includegraphics[width=30pt,height=37.5pt]{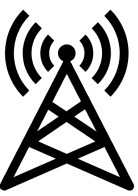}};
\draw (331.21,83.39) node  {\includegraphics[width=28.82pt,height=26.08pt]{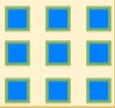}};
\draw (451.21,178.89) node  {\includegraphics[width=12.32pt,height=16.33pt]{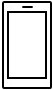}};
\draw (452.71,76.89) node  {\includegraphics[width=26.57pt,height=25.33pt]{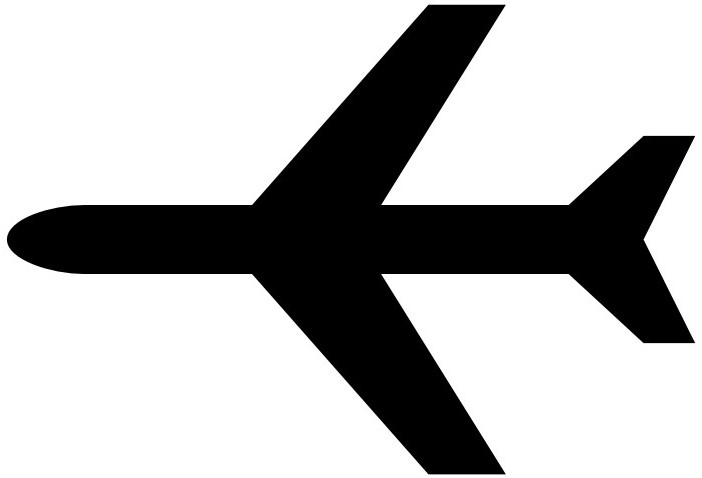}};
\draw [color={rgb, 255:red, 74; green, 144; blue, 226 }  ,draw opacity=1 ]   (266,192) -- (331.43,99.78) -- (430.89,182.5) ;
\draw [shift={(432.43,183.78)}, rotate = 219.75] [color={rgb, 255:red, 74; green, 144; blue, 226 }  ,draw opacity=1 ][line width=0.75]    (10.93,-3.29) .. controls (6.95,-1.4) and (3.31,-0.3) .. (0,0) .. controls (3.31,0.3) and (6.95,1.4) .. (10.93,3.29)   ;
\draw [color={rgb, 255:red, 74; green, 144; blue, 226 }  ,draw opacity=1 ]   (265,206) -- (422.43,204.79) ;
\draw [shift={(424.43,204.78)}, rotate = 179.56] [color={rgb, 255:red, 74; green, 144; blue, 226 }  ,draw opacity=1 ][line width=0.75]    (10.93,-3.29) .. controls (6.95,-1.4) and (3.31,-0.3) .. (0,0) .. controls (3.31,0.3) and (6.95,1.4) .. (10.93,3.29)   ;
\draw [color={rgb, 255:red, 126; green, 211; blue, 33 }  ,draw opacity=1 ]   (261.31,180.83) -- (331.43,99.78) -- (429.45,86.05) ;
\draw [shift={(431.43,85.78)}, rotate = 172.03] [color={rgb, 255:red, 126; green, 211; blue, 33 }  ,draw opacity=1 ][line width=0.75]    (10.93,-3.29) .. controls (6.95,-1.4) and (3.31,-0.3) .. (0,0) .. controls (3.31,0.3) and (6.95,1.4) .. (10.93,3.29)   ;
\draw [shift={(260,182.34)}, rotate = 310.86] [color={rgb, 255:red, 126; green, 211; blue, 33 }  ,draw opacity=1 ][line width=0.75]    (10.93,-3.29) .. controls (6.95,-1.4) and (3.31,-0.3) .. (0,0) .. controls (3.31,0.3) and (6.95,1.4) .. (10.93,3.29)   ;
\draw [color={rgb, 255:red, 255; green, 0; blue, 0 }  ,draw opacity=1 ] [dash pattern={on 0.84pt off 2.51pt}]  (410.53,116.5) -- (404.43,119.78) -- (440.9,96.4) ;
\draw [shift={(443.43,94.78)}, rotate = 147.34] [fill={rgb, 255:red, 255; green, 0; blue, 0 }  ,fill opacity=1 ][line width=0.08]  [draw opacity=0] (8.04,-3.86) -- (0,0) -- (8.04,3.86) -- (5.34,0) -- cycle    ;
\draw (463.14,205.17) node  {\includegraphics[width=13.07pt,height=17.08pt]{Images/user.jpg}};
\draw (437.71,205.39) node  {\includegraphics[width=13.07pt,height=17.08pt]{Images/user.jpg}};
\draw (224,221) node [anchor=north west][inner sep=0.75pt]   [align=left] {\textbf{{\footnotesize {\fontfamily{ptm}\selectfont DFBS}}}};
\draw (385,221) node [anchor=north west][inner sep=0.75pt]   [align=left] {\textbf{{\footnotesize {\fontfamily{ptm}\selectfont Communications users}}}};
\draw (443.43,94.78) node [anchor=north west][inner sep=0.75pt]   [align=left] {\textbf{{\footnotesize {\fontfamily{ptm}\selectfont Target}}}};
\draw (310,51) node [anchor=north west][inner sep=0.75pt]   [align=left] {{\footnotesize {\fontfamily{ptm}\selectfont \textbf{BD-RIS}}}};
\draw (340,155) node [anchor=north west][inner sep=0.75pt]  [font=\scriptsize] [align=left] {{\scriptsize \textcolor[rgb]{1,0.01,0.01}{\textbf{Blocked LoS}}}};
\end{tikzpicture}}